\documentclass[aip,amsmath,reprint]{revtex4-2}

\usepackage{graphicx}
\usepackage{dcolumn}
\usepackage[mathlines]{lineno}

\usepackage[utf8]{inputenc}
\usepackage[T1]{fontenc}

\newcommand{\subfigimg}[4][,]{%
  \setbox1=\hbox{\includegraphics[#1]{#3}}
  \leavevmode\rlap{\usebox1}
  \rlap{\hspace*{#4}\raisebox{\dimexpr\ht1-2\baselineskip/2}{#2}}
  \phantom{\usebox1}
}
\newcommand{\reffig}[1]{Fig.\,\ref{#1}}
\newcommand{\degree}{\(^\circ\)}
\newcolumntype{P}[1]{>{\centering\arraybackslash}p{#1}}
\newcommand{\citeintext}[2]{[\kern-#2em\citenum{#1}]}

\draft 
\begin{document}

\title{Electron Spectroscopy using Transition-Edge Sensors}

\author{K. M. Patel}
\affiliation{Cavendish Laboratory, University of Cambridge, 
JJ Thomson Avenue, Cambridge CB3 0HE, United Kingdom}
\affiliation{National Physical Laboratory, Hampton Road, Teddington TW11 0LW, United Kingdom}

\author{S. Withington}
\affiliation{Department of Physics, University of Oxford, Oxford OX1 3PU, United Kingdom}

\author{A. G .Shard}
\affiliation{National Physical Laboratory, Hampton Road, Teddington TW11 0LW, United Kingdom}

\author{D. J. Goldie}
\affiliation{Cavendish Laboratory, University of Cambridge, 
JJ Thomson Avenue, Cambridge CB3 0HE, United Kingdom}

\author{C. N. Thomas}
\affiliation{Cavendish Laboratory, University of Cambridge, 
JJ Thomson Avenue, Cambridge CB3 0HE, United Kingdom}

\date{\today}
\begin{abstract}
Transition-edge sensors (TESs) have the potential to perform electron spectroscopic  measurements with far greater measurement rates and efficiencies than can be achieved using existing electron spectrometers. Existing spectrometers filter electrons by energy before detecting a narrow energy band at a time, discarding the vast majority of electrons available for measurement. In contrast, transition-edge sensors (TES) have intrinsic energy sensitivity and so do not require prior filtering to perform energy-resolved measurements. Despite this fundamental advantage, TES electron spectroscopy has not, to our knowledge, previously been reported in the literature. We present the results of a set of proof-of-principle experiments demonstrating TES electron spectroscopy experiments using Mo/Au TESs repurposed for electron calorimetry. Using these detectors, we successfully measured the electron spectrum generated by an electron beam striking a graphite target with energies between 750 and 2000\,eV, at a noise-limited energy resolution of 4\,eV. Based on the findings of these experiments, we suggest improvements that could be made to TES design to enhance their electron detection capabilities through the use of of a dedicated electron absorber in the device with integrated electron optics.
\end{abstract}

\maketitle
\newpage

\section{\label{sec:Intro}Introduction}
Transition-edge sensors (TESs) are thin-film superconducting devices capable of high-sensitivity, energy-resolved photon measurement. Over the last thirty years, TESs have found applications across an increasing range of fields from astronomic observations to dark matter searches \cite{ullomReviewSuperconductingTransitionedge2015a,smithPerformanceBroadBandHighResolution2021,sobrinDesignIntegratedPerformance2022a,hazumiLiteBIRDSatelliteStudies2019,straussPrototypeDetectorCRESSTIII2017,rauSuperCDMSSNOLABStatus2020}. However, one area of TES research that has received little attention is massive particle spectroscopy, encompassing molecular, ion-beam and electron measurement techniques. The lack of research into TES electron spectroscopy is of particular note due to the widespread usage of electron spectroscopic techniques and the potential benefits offered by TESs over conventional electron spectrometers.


All modern commercial electron spectrometers follow the same fundamental operating principle where electrons are collected, dispersed by energy and then counted using energy-insensitive detectors. For example, the concentric hemispherical analyser (CHA), the analyser of choice for X-ray photoelectron spectroscopy (XPS) measurements, uses the electric field between two concentric hemispheres of differing electric potentials to disperse electrons in space depending on their energy. This arrangement sets up an energy filter where only electrons within a narrow energy band, defined by the hemisphere potentials, can pass through the hemispheres to the particle-counting microchannel plate.

The inefficiency in this form of measurement lies in the fact that only a narrow electron energy band, approximately 1 to 10\,eV in width, can be measured at a time. To perform a wide spectrum measurement, this narrow band must be swept across the entire spectral range and so, at every individual moment in time, the vast majority of electrons emitted from the sample and collected by the instrument cannot be measured. If an energy range of 1\,keV is to be measured, a CHA measuring a window of 1\,eV would have at best a measurement efficiency of 0.1\,\% across the measurement. The effect of this inefficiency can be mitigated by increasing the number of electrons collected, either by emitting more electrons from the sample or widening the solid angle of collection; however, these measures cannot address the underlying inefficiency within the operating principle of the spectrometer itself.

An alternate approach is time-of-flight electron spectrometry\cite{DeFanis}. In this case, a pulsed X-ray source is required and this is only achievable with specialized and expensive equipment such as an X-ray Free Electron Laser (XFEL). The operation requires fast detection electronics and the electron time-of-flight can be converted into kinetic energy, which scales as the inverse square of the flight time. In principle, all emitted electrons entering the analyzer can be detected, but with variable energy resolution which depends upon the pulse width of the X-ray source, the detector electronics and the time of flight. Only low-energy electrons are detected with good energy resolution and thus the electrons are typically retarded using an electric field prior to entering the analyzer. For practical purposes, the efficiency, resulting from the duty cycle of X-rays and the necessity to collect multiple kinetic energy regions, is low.

The ideal solution is to use a detector that is intrinsically able to resolve the energy of an incident electron, removing the need for filtering in space or time. Such a detector would be capable of continuous measurement and the instantaneous fraction of electrons emitted from the sample that can be characterized will scale with the number of detectors used. TESs are near-ideal candidate, as they have high energy sensitivities and the technology exists to readout out arrays of several thousand devices concurrently \cite{smith100000PixelMicrocalorimeter2020,stevensCharacterizationTransitionEdge2020,gottardiReviewXrayMicrocalorimeters2021}.

TESs perform high-resolution particle calorimetry by exploiting the extremely sharp resistance-temperature dependence of certain metals in their superconducting transition. Voltage-biasing a TES at a point within its superconducting transition creates an electrothermal feedback loop that maintains the device's temperature by balancing the thermal power received from particle absorption with a reduction in ohmic power dissipated. The TES current drop associated to the drop in ohmic power dissipation can be monitored to precisely determine the energy of the absorbed particle.

An additional advantage of TES measurement over existing electron analysers is the ability of every pixel to independently measure both electron arrival time and energy simultaneously. This ability allows for coincidence spectroscopy measurements, where coincident electron emission events are analysed to probe energy transfer mechanisms within the sample\cite{arionCoincidenceSpectroscopyPresent2015}; these measurements could be performed alongside conventional electron spectroscopic techniques without requiring additional, specialised apparatus.


In a previous article, we simulated the capabilities of a TES array measuring an X-ray photoelectron spectrum\cite{patelSimulationMethodInvestigating2021a}, the results of which showed that small arrays of TESs, numbering tens of devices, could be capable of comparable energy resolutions and measurement rates to existing XPS analysers, with considerable scope of improving measurement rates by increasing array size. Based upon these simulations, TES electron spectroscopy is a realistic proposal but experimental validation of TES electron spectroscopy has not previously been provided.
This paper reports the first experimental demonstration of TES electron spectroscopy.

\section{\label{sec:Exp}Experimental}
\begin{figure}
\begin{tabular}{@{}P{0.95\linewidth}}
    \subfigimg[width = \linewidth]{}{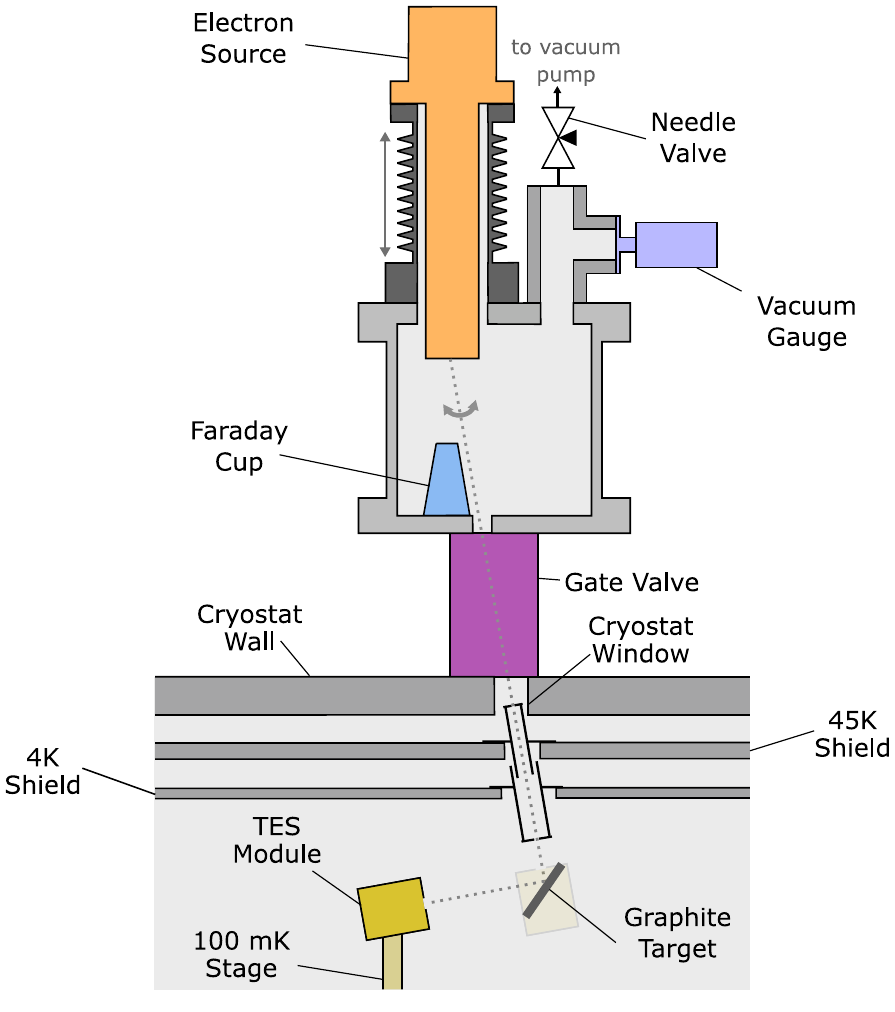}{-12pt}
\end{tabular}
\caption{TES electron measurement experiment in the scattered electron measurement beam configuration.}
\label{fig:ESourceDiagram}
\end{figure}
Proof-of-principle TES electron measurements were performed using Mo/Au TESs that were adapted for the purposes of electron calorimetry. The Mo/Au TES used consisted of a square superconducting bilayer, 70\,\(\mu\)m in length, with film thicknesses of 120\,nm gold atop 40\,nm of molybdenum; this bilayer was suspended by four 1.41\,\(\mu\)m long Si\(_x\)N\(_y\) legs, and displayed a transition temperature (\(T_c\)) of 200\,mK. Further details on the device design and fabrication have been reported previously \cite{harwinMicroscopicPhysicsTransition2021}. 

A SPECS EQ22 electron source was mounted on a closed-cycle adiabatic demagnetisation refrigerator in the manner shown in \reffig{fig:ESourceDiagram}. The electron source was used to generate electron beams with energies ranging from 250 to 2000\,eV with the position of the beam controlled by X- and Y- electrostatic deflector plates within the source itself. The electron beam was directed into the cryogenic volume through two tubes mounted to the cryostat heat shields and capped with apertures.

The TES device chip was mounted to the 100\,mK stage of the cryostat and read-out with an amplifier that uses superconducting quantum interference devices (SQUIDs). To prevent charge accumulation, the TES was connected to a shared cryostat ground through the TES bias circuit. 

Two experimental configurations were tested which will be referred to as direct and scattered measurements. The direct measurements were performed by positioning the TES module, containing the TES and SQUID arrays, in front of the cryostat window, receiving the electron beam directly. The purpose of these direct measurements were to determine beam alignment through the cryostat windows, demonstrate TES electron detection and characterise the response of the TES to differing electron beam energies. For the scattered measurements, the electron beam was aimed at a target of 0.254\, mm thick graphite foil (99.8\% purity) at an incidence angle of 60\degree, with the TES  measuring the scattered electron spectrum.

A key consideration in the experimental design was mitigating infrared black-body radiation emitted from room-temperature components onto the TES module. No suitable window material exists that allows unobstructed electron passage through it whilst filtering infrared radiation at the electron energies being tested, a free space path was used from the source to the TES. Infrared loading was reduced along this path using a combination of four approaches.

Firstly, the electron beam was deflected off-axis into the cryostat, as shown in \reffig{fig:ESourceDiagram}; this deflection removed line-of-sight from the high-temperature electron source filament into the cryostat.

The electrons passed through two tubes mounted to the 45\,K and 4\,K stages respectively before entering the cryostat chamber. These tubes were painted black with colloidal graphite paint to absorb infrared radiation whilst providing an electrically grounded surface to prevent charge accumulation.

The third modification was to place apertures on the TES module (200\,\(\mu\)m diameter), 4\,K tube (3\,mm diamater for direct measurements, 1\,mm diameter for scattered measurements) and 45\,K tube (5\(\times\)3\,mm slotted aperture), to block stray infrared photons reaching the detector. 

The final infrared mitigation was the use of mesh grid across the 45\,K aperture to diffract and screen incident radiation. A mesh grid was also used on the 4\,K aperture for the direct measurements. The meshes consisted of 1500 lines per inch copper grids with 55\% open area and approximately 10\,\(\mu\)m hole width. The presence of the mesh grids was a compromise between greatly reducing infrared loading and blocking or scattering a portion of the electron beam entering the cryostat.

The experiment was designed to measure electrons with energies from 0 to 2000\,eV; at these energies, electrons are highly susceptible to deflection by the Earth's magnetic field. To reduce the impact of this deflection on the measurements, the electron source was shielded with mu-metal and Metglas 2705M magnetic shielding foil was wrapped around the electron source vacuum chamber and infrared tubes. The aperture on the 45\,K tube was not circular but horizontally slotted with dimensions of 5\(\times\)3\,mm to compensate for magnetic beam deflection prior to entering the cryostat.
\section{\label{sec:Results}Results}

\subsection{\label{sec:Direct}Direct Measurements}
\begin{figure}
\centering
\begin{tabular}{@{}P{0.99\linewidth}}
    \subfigimg[width = \linewidth]{}{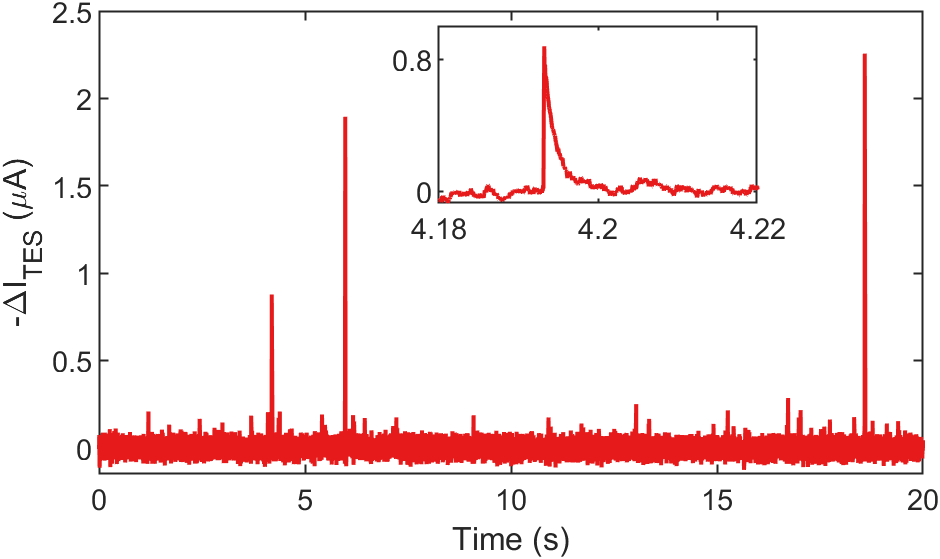}{-12pt}
\end{tabular}
\caption{Time series TES measurement of a 2\,keV electron beam. The inset shows a measurement event with estimated 190\,eV energy.}
\label{fig:D1_TS}
\end{figure}
The response of a TES to an incident particle is an exponentially decaying pulse in current whose area directly relates to the energy absorbed by the device. \reffig{fig:D1_TS} shows a time series TES measurement of a 2\,keV electron beam aimed at the TES module in which a series of such pulses were observed.

Measurements such as that shown in \reffig{fig:D1_TS} were analysed to determine the energy spectrum of observed electron events. The analysis software used to identify, extract and calculate the energies of individual TES events was adapted from a previous work simulating TES electron spectroscopy\cite{patelSimulationMethodInvestigating2021a}. The energies of TES calorimetry peaks would typically be determined by comparison to a known energy standard, such as X-ray emission lines in the case of TES X-ray calorimetry. For these electron measurements, in the absence of such a calibration standard, the absorbed particle energies \( E_\text{TES}\) have been calculated by \cite{irwinTransitionEdgeSensors2005}
\begin{equation}\label{eq1}
     E_\text{TES} =  \int_{t_1}^{t_2} -\Delta I_\text{TES}(t) V_\text{TES}\,dt, 
\end{equation}
where \(t_1\) and \(t_2\) are the start and end times of the electron absorption event, \(V_\text{TES}\) is the TES bias voltage and \(I_\text{TES}\) is the change in TES current from equilibrium. The calculated energy corresponds to the integrated change in the Joule heating within the TES due to an electron absorption event. This energy is equivalent to energy absorbed by the TES in the limit of strong electrothermal feedback where all of the received energy is compensated for electrically; in practice, a portion of the received thermal energy instead diffuses to the bath. The energy difference can be approximated by \cite{figueroa-felicianoTheoryDevelopmentPositionsensitive2001}
\begin{equation}\label{eq2}
     E_\text{TES} =  \frac{\left(1- \frac{T_b^n}{T_c^n}\right)}{\left(1- \frac{T_b^n}{T_c^n}\right)+\frac{n}{\alpha}
     }E_\text{abs} 
\end{equation}
where \(T_b\) is the bath temperature, \(n\) is a device parameter with a value between 2 and 4 that is characteristic of the thermal link between the TES and surrounding thermal bath, and \(\alpha\) is a measure of the sharpness of the TES superconducting transition at the device temperature. For example, a device with  \(\alpha = 30\), \(n = 2\),  \(T_b= 130\)\,mK and  \(T_c = 200\)\,mK would lose 10\% of the received energy by thermal diffusion. It should be noted that \(\alpha\) is a function of temperature and will vary within a pulse as the TES temperature moves within the superconducting transition, which further complicates the calculation. The presence of this energy underestimate is not significant for the purposes of these proof-of-principle experiments because it is systematic to all electron observations and is caused by the analysis method used rather than the intrinsic behaviour of the device.

\begin{figure}
\centering
\begin{tabular}{@{}P{0.95\linewidth}}
    \subfigimg[width = \linewidth]{a)}{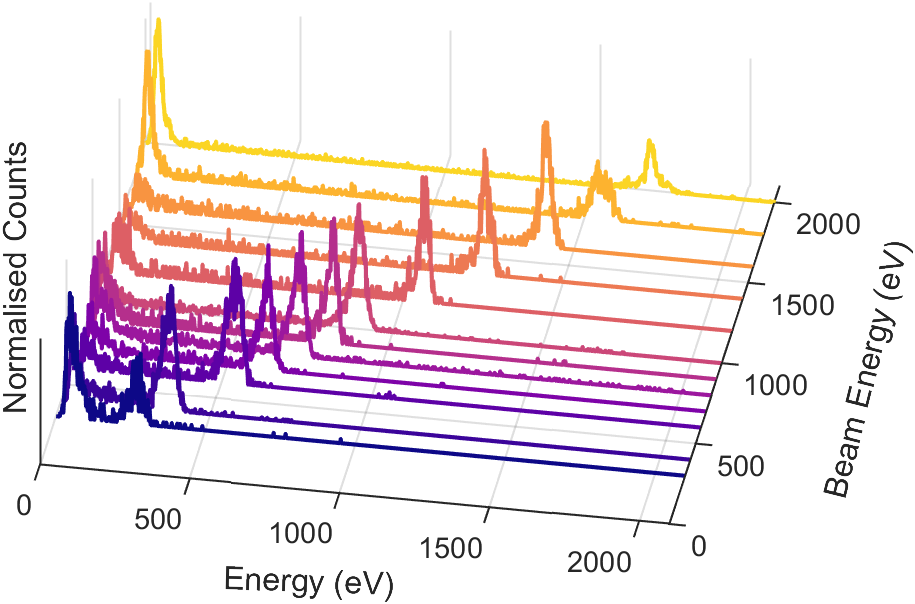}{-10pt}
\end{tabular}
\begin{tabular}{@{}P{\linewidth}}
    \subfigimg[width = 0.95\linewidth]{b)}{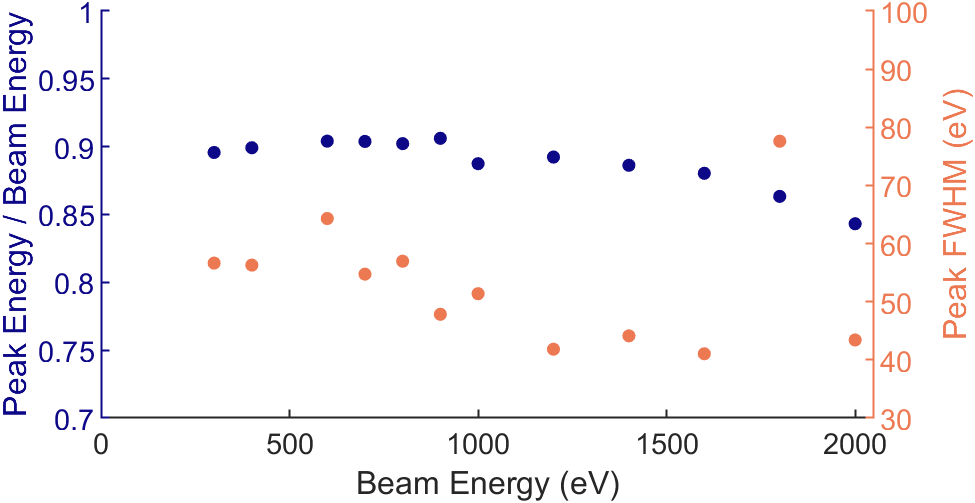}{-12pt}
\end{tabular}
\caption{a) Measured electron spectra taken at beam energies spanning 300\,eV to 2000\,eV. The spectra have been normalised by total area and energy bin widths of 2\,eV have been used. b) Ratio between measured energy of the high energy peak and the applied beam energy plotted on the left axis. The measured peak energy was calculated by applying a Gaussian fit. Full-width at half maximum of the fitted high-energy peak plotted against beam energy on the right axis.}
\label{fig:D2_Spec}
\end{figure}

The measurement in \reffig{fig:D1_TS} was repeated across multiple beam energies ranging from 250\,eV to 2000\,eV; the resulting measured electron energy distributions are shown in \reffig{fig:D2_Spec}a. Several common features are seen across all energies: a high energy peak that tracks the beam energy, a fixed low energy peak at 20\,eV and a flat background of events. The high energy peak very likely corresponds to primary electrons from the electron source, with the low energy events being secondary electrons emitted from the meshes used to reduce thermal loading or other surfaces between the electron source and the TES itself. The low-level background of events is likely a result of inelastic scattering during flight, or upon striking the TES. In either case, this scattering would result in partial energy absorption by the TES. A small number of events with energies in excess of the beam energy were observed; these result when two or more electrons are detected near simultaneously and are resolved by the analysis software as a single event, with energy equal to the sum of their individual energies. This effect, termed pile-up, is common to other detectors.

The primary electron peaks were fitted with Gaussian distributions to identify the peak location relative to the beam energy, and the peak full-width at half maximum (FWHM) (\reffig{fig:D2_Spec}b). The ratio of central peak energy to beam energy is approximately 0.9 up to 1000\,eV and then falls at greater beam energies. This trend is consistent with the expected bias from the energy analysis method. The energy received by electron absorption briefly increases the TES temperature before being compensated for by the electrothermal feedback loop. For high energy particles, the TES temperature can rise through a significant portion of its superconducting transition or be driven into its normal state. As the TES reaches the upper region of its transition, \(\alpha\) falls in value, reducing the effectiveness of the electrothermal feedback loop and increasing the energy underestimate.

\begin{figure}
\centering
\begin{tabular}{@{}P{0.95\linewidth}}
    \subfigimg[width = 0.9\linewidth]{}{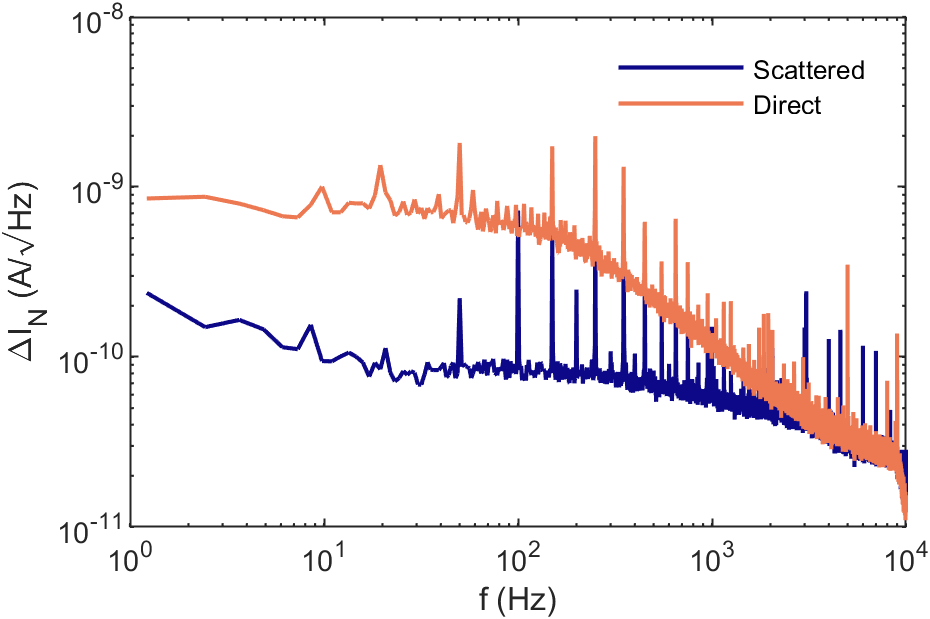}{-12pt}
\end{tabular}
\caption{TES current noise spectral densities for direct beam and scattered measurements.}
\label{fig:TES_PSD}
\end{figure}

The broad FWHM of the primary electron peak is primarily due to the energy resolution of the TES. A disadvantage of the direct beam configuration is infrared loading from the room-temperature apparatus onto the TESs with the corresponding photon noise significantly degrading energy sensitivity. This degradation is shown by the current-noise spectral density measurements in \reffig{fig:TES_PSD}. The TES FWHM energy resolution can be approximated by \cite{moseleyThermalDetectorsRay1984}
\begin{equation}\label{eq3}
    \Delta E_{abs} = 2\sqrt{2 \text{log}(2)}\,\text{NEP}
    \left(0\right)\sqrt{\tau_\text{eff}},
\end{equation}
where NEP\((0)\) is the zero-frequency noise-equivalent power and \(\tau_\text{eff}\) is the TES effective response time. NEP\((0)\) can be calculated using the zero-frequency TES responsivity \(s_I(0)\), following \cite{irwinTransitionEdgeSensors2005}
\begin{equation}\label{eq4}
   \text{NEP}(0) = \frac{\Delta I_N(0)}{s_I(0)}
\end{equation}
 where \(\Delta I_N\) is the current-noise spectral density. With  \(\tau_\text{eff} = 1.1\)\,ms and \(s_I(0) = 17\) and 21\,\(\mu\)A/pW for the direct and scattered measurements respectively, the predicted resolutions are calculated as 20\,eV for the direct measurements compared to 4\,eV for the scattered measurements using the same device; this change in resolution can be attributed to a reduction in thermal loading and corresponding photon noise in the scattered measurements.

\subsection{\label{sec:Scattered}Scattered Measurements}
\begin{figure}
\centering
\begin{tabular}{@{}P{0.48\linewidth}@{}P{0.01\linewidth}@{}P{0.48\linewidth}}
    \subfigimg[width = \linewidth]{a)}{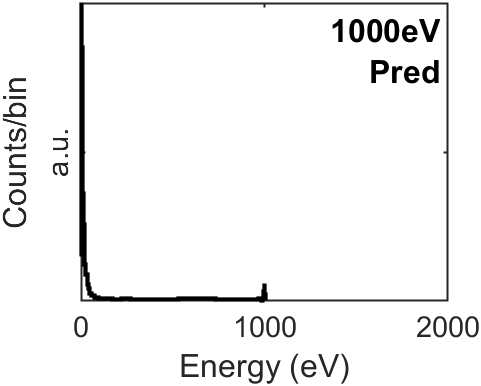}{-0pt}&&
    \subfigimg[width = \linewidth]{b)}{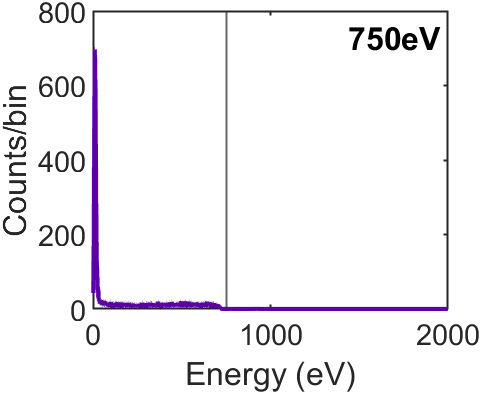}{-0pt}\\
    \subfigimg[width = \linewidth]{c)}{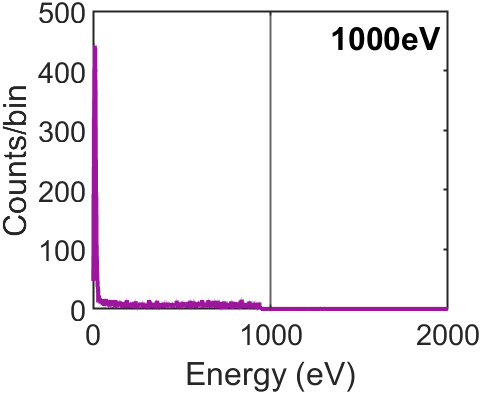}{-0pt} &&
    \subfigimg[width = \linewidth]{d)}{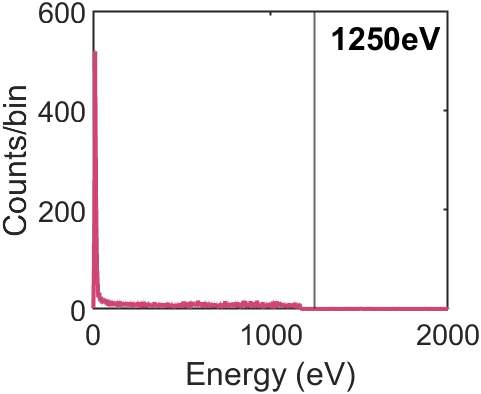}{-0pt}\\
    \subfigimg[width = \linewidth]{e)}{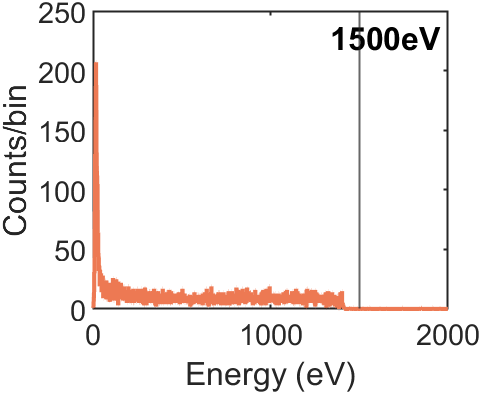}{-0pt} &&
    \subfigimg[width = \linewidth]{f)}{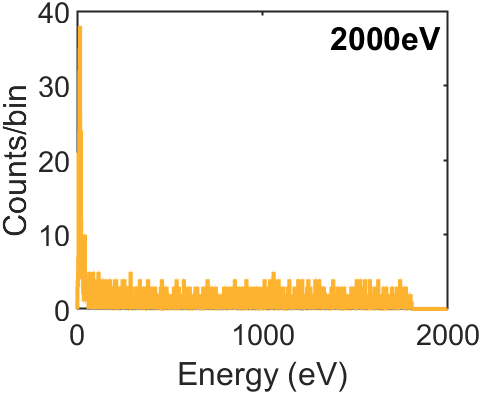}{-0pt}
\end{tabular}
\caption{Measured electron energy spectra generated by electron impact on a graphite target with a 1\,eV bin width. The incident electron beam energy on the target is shown by the vertical line.  The predicted spectrum in (a) has been reproduced from measurements by Goto and Takeichi \cite{gotoCarbonReferenceAuger1996a} for comparison to observed TES spectra.}
\label{fig:G3_Spec}
\end{figure}

The scattered electron measurements show the use of a TES in a spectroscopic role measuring the scattered electron spectrum emitted from a graphite target. The form of this spectrum can be predicted using the measurements by Goto and Takeichi \cite{gotoCarbonReferenceAuger1996a} where such an electron spectrum was generated by a 1\,keV beam and measured using a cylindrical mirror analyser (\reffig{fig:G3_Spec}a).
The TES measured electron spectra are shown in \reffig{fig:G3_Spec}b\,-f at beam energies from 750\,eV to 2000\,eV. These measurements agree well with the comparison spectrum in \reffig{fig:G3_Spec}a, showing a sharp secondary electron peak below 50\,eV and a background of inelastically backscattered electrons extending up to, but not exceeding the beam energy. It is important to highlight that the spectra in \reffig{fig:D2_Spec}a and \reffig{fig:G3_Spec} were constructed by aggregating the energies of individual electron detection events. As such, the low-level fluctuations in the observed spectra reflect the number of electrons observed in the relevant energy bin and are not background noise.

The observed spectra do not show the elastic peak, likely due to the peak being broadened into the background as a result of electron absorption inefficiency of the TES, the energy resolution of the detector and analysis method used. In these devices, which were not designed for electron detection, the electrons were absorbed directly in the Mo/Au TES. The gold surface of the device will have emitted a significant number of secondary electrons. Based on previous measurements \cite{patelTransitionEdgeSensorsElectron2023}, we estimate a secondary electron yield of 1.4 from the gold film per primary electron at 1\,keV beam energy with the vast majority of secondary electrons with energies between 0 and 20\,eV. In a device designed specifically for electron detection, this effect can be greatly mitigated by using a separate absorber made from a more favorable material, such as carbon.

\begin{figure}
\centering
\begin{tabular}{@{}P{0.95\linewidth}}
    \subfigimg[width = 0.8\linewidth]{}{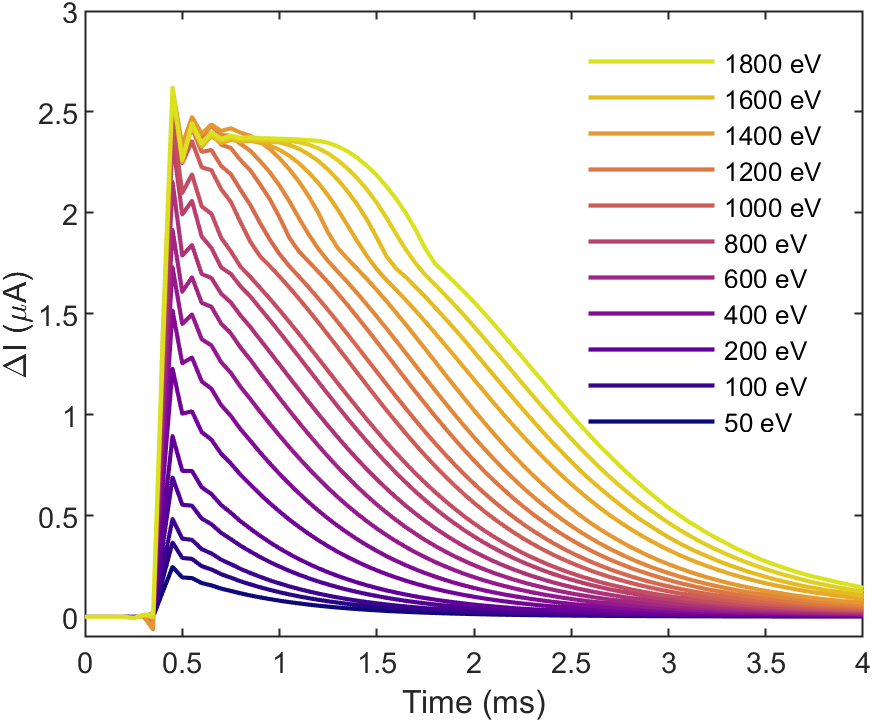}{-6pt}
\end{tabular}
\caption{Average TES responses based on calculated event energy. Each plotted response has been averaged from observed responses within 5\,eV of the nominal event energies. Responses between 200\,eV and 1800\,eV event energies have been plotted at 100\,eV intervals; the legend only shows events in this range at 200\,eV intervals for concision.}
\label{fig:G3_Pulses}
\end{figure}

The TES response to electron events of different energies is shown in \reffig{fig:G3_Pulses} where TES pulse shapes have been grouped and averaged by energy. The figure clearly demonstrates the progression of the TES response into saturation as electron energy increases. The TES is said to saturate when the absorbed energy drives the device into its normal state where the devices loses sensitivity to additional energy input, and the TES response plateaus. This effect can be seen at electron energies at 1400\,eV and above. While the TES loses sensitivity to additional energy input under these conditions, the electrothermal feedback loop continues to compensate for the thermal energy within the device leading to a lengthened TES response and the particle energy can still be estimated, as demonstrated in \reffig{fig:G3_Spec}f.

\section{\label{sec:Discussion}Discussion}
The measurements in \reffig{fig:G3_Spec} experimentally demonstrate the principle of TES electron spectroscopy and highlight the most important areas of development required to improve detector performance. The advantage of TES electron spectroscopy over existing methods is potential orders of magnitude improvement to electron measurement efficiency. However, to be of practical use, the TES must provide energy resolutions within the ranges achievable by existing electron spectrometers.

In the case of XPS measurements, the resolution of an XPS analyser typically lies in the range of 0.1-5\,eV for electrons below 1500\,eV. For comparison, phonon-noise-limited TESs with a transition temperature of 200\,mK are capable of energy resolutions better than 1\,eV. The noise-limited TES resolution scales with temperature, with lower transition temperatures providing improved resolution. As such, within the bounds of current cryostat capabilities, a practical TES spectrometer for XPS would be operating near its phonon-noise-limited resolution and so should be optimised for energy resolution.

The resolution of a TES is determined by the efficiency of the device at capturing the energy of the absorbed particle, and the ability to extract the magnitude of absorbed energy from the measured detector response. The measurements in the previous section show no apparent differences in the behaviour of the TES after electron absorption compared to TES behaviour in photon calorimetry. Sub-eV resolutions have been demonstrated using TES X-ray calorimeters occupying a similar energy range to that investigated here \cite{leeFinePitchTransitionedge2015a}, indicating that such TES performance in electron calorimetry is entirely reasonable. However, achieving such resolutions would require near ideal electron energy absorption efficiencies.

The main energy loss pathways during electron absorption are backscattered electron emission and secondary electron emission, examples of which can be clearly seen in \reffig{fig:G3_Spec} where such electrons scattered from a graphite surface were measured. Backscattered electrons can carry a wide range of energies up to the primary incident electron energy whereas nearly all secondary electrons have energies below 50\,eV \cite{dingEnergySpectraBackscattered2004a,walkerSecondaryElectronEmission2008}. In a TES electron spectroscopy measurement, the effect of backscattered emission is to map received electron energies across a range of energies throughout the spectrum, distorting the background. Secondary electrons shift observed electron energies by relatively small amounts, distorting the characteristic spectral peaks. In both cases, the rate of emission from the TES absorber can be reduced by using absorbing material made of low atomic mass materials due to their lower inelastic scattering cross-sections \cite{albertc.thompsonXrayDataBooklet2009}. In addition, using roughened or structured (e.g. pitted) absorber surfaces can reduce electron emission.

A promising method to greatly improve electron absorption efficiencies is through the use of electron optics. For example, placing a surface above a TES (with an aperture for electron transmission), biased at -20\,V, would suppress the majority of secondary electron emitted from the TES surface. The use of electron optics to enhance TES electron absorption raises the question of possible electric field coupling to the TES and degradation in performance. Measurements of the behaviour of a TES in the presence of DC electric fields up to 90\,kV/m showed no observable impact on TES behaviour, indicating that electrostatic optics can be practically implemented in a TES electron spectrometer \cite{patelSensitivityTransitionedgeSensors2023}.

\begin{figure}[t]
\centering
\begin{tabular}{@{}P{0.95\linewidth}}
    \subfigimg[width = \linewidth]{}{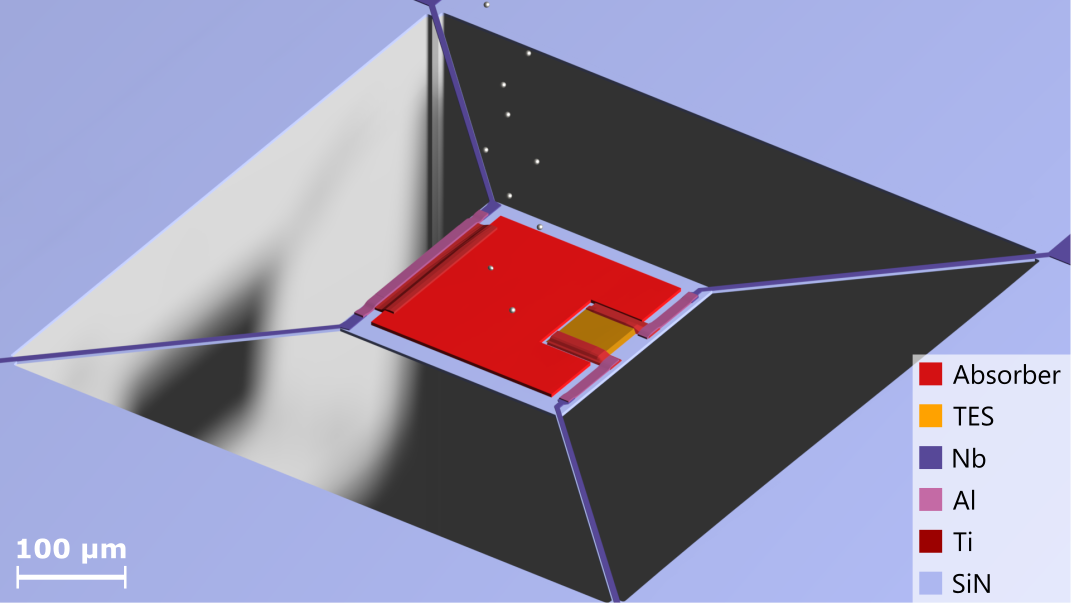}{-6pt}   
\end{tabular}
\caption{Proposed design of a bespoke electron spectroscopy TES with an independently biased absorber.}
\label{fig:ETES}
\end{figure}

An alternative approach could be to integrate electron micro-optics into the TES itself. A TES design based upon this approach is shown in \reffig{fig:ETES} using a Ti/Au bilayer superconductor and a dedicated Au/Ti electron absorber, with the surface layer of the absorber being titanium. In this design, the absorber is electrically isolated from the TES allowing it to be biased independently; applying a +20\,V DC bias to this absorber would have an equivalent effect as applying a -20\,V bias to an external surface as described previously. In this way, it is possible to achieve highly-efficient electron absorption independent of absorber material choice.

There also exists a range of possibilities with regards to the manner in which a TES can be used in an electron spectrometer. The TESs could perform the entirety of the electron energy discrimination, as was the case in \reffig{fig:G3_Spec} but there may be advantages in a hybrid approach, combining TES calorimetry with electron energy dispersion used in existing electron spectrometers. TES energy resolution follows \(\Delta E \propto \sqrt{E_\text{sat}}\) where \(E_\text{sat}\) is the TES saturation energy, and so TESs with lower saturation energy display improved resolution. If the electrons collected by a spectrometer are spatially dispersed by energy and separated into bins, the electrons can be uniformly decelerated by a known amount within each bin independently, before reaching the TES detector array. Controllable electron deceleration prior to detection would allow for TESs with saturation energies far below the maximum of the emitted electron spectrum, greatly improving the available energy resolution and the presence of multiple bins would allow for concurrent measurement of the entire spectral range. Larger number of bins can be used with reduced saturation energies and improved resolutions, with the possibility of exceeding the energy resolution, detection efficiency and count rate capabilities of existing electron spectrometers simultaneously.

\section{Conclusion}
Due to their fundamentally different method of performing energy-resolved measurements, TESs offer significant benefits in electron measurement over existing electron spectrometers. The inherent energy sensitivity of TESs allows for orders of magnitude improvement in electron detection efficiency, and therefore measurement rate, over what is currently achievable. We believe the presented measurements experimentally demonstrate TES electron spectroscopic measurement, opening the door to further investigation in this field. The ability to perform spectroscopic measurements using TESs that have been adapted for electron calorimetry suggests far greater performance is achievable using devices designed specifically for electron measurement and that creating a TES array capable of exceeding the capabilities of traditional electron spectrometers is entirely possible.

The experimental work has been performed in the context of electron spectroscopy but applies more widely to TES massive particle spectroscopy in general, an area that has received little attention in comparison to TES photon measurement. The ability to use TESs for the measurement of charged particles is of particular interest as the use of electron optics allows for precise control of the acceleration and position of these particles. Integrating electron optical systems with TES spectrometers can allow for manipulating the interaction between the particle and the TES, enhancing or reducing absorption efficiency, modifying the energy of the incident particles or screening them entirely. The ability to screen charged particles is also of interest in applications beyond massive particle spectroscopy, such as in space-based astronomical observation where secondary electrons produced by cosmic ray strikes can result in unwanted measurement events in the detector. An elegant route control the interaction between TESs and charged particles is to integrate electron optics into the detector itself. The proposed TES electron calorimeter design uses an independently biased absorber to enhance electron energy absorption efficiency but such a structure can equally be used to screen low energy electrons or indicating the scope of possibility in this approach to particle detection.

\begin{acknowledgments}
KMP acknowledges support from EPSRC Cambridge NanoDTC (EP/L015978/1). KMP and AGS acknowledge support from the National Measurement System of the UK Department for Science, Innovation and Technology. We would also like to thank Rebecca Harwin and Michael Crane for their work on the initial device designs and fabrication, as well as David Sawford and Dennis Molloy for their assistance in preparing the experiments.
\end{acknowledgments}

\section*{Author Declarations}
\subsection*{Conflict of Interest}
The authors have no conflicts to disclose.

\subsection*{Author Contributions}
\textbf{K. M. Patel}: Conceptualization, Methodology, Investigation (lead), Software, Formal analysis (lead), Writing - original draft, Visualization.
\textbf{S. Withington}: Conceptualization, Methodology, Supervision, Project administration, Funding acquisition, Writing – review and editing.
\textbf{A. G. Shard}: Conceptualisation, Methodology, Supervision, Funding acquisition, Writing – review and editing.
\textbf{D. J. Goldie}: Conceptualization, Investigation (supporting), Formal analysis (supporting), Writing – review and editing.
\textbf{C. N. Thomas}: Conceptualization, Formal analysis (supporting), Resources, Writing – review and editing.

\section*{Data Availability Statement}
 The data that support the findings of this study are openly available in the Apollo repository at http://doi.org/10.17863/CAM.106426, reference number.

\bibliography{main.bib}

\begin{thebibliography}{23}%
\makeatletter
\providecommand \@ifxundefined [1]{%
 \@ifx{#1\undefined}
}%
\providecommand \@ifnum [1]{%
 \ifnum #1\expandafter \@firstoftwo
 \else \expandafter \@secondoftwo
 \fi
}%
\providecommand \@ifx [1]{%
 \ifx #1\expandafter \@firstoftwo
 \else \expandafter \@secondoftwo
 \fi
}%
\providecommand \natexlab [1]{#1}%
\providecommand \enquote  [1]{``#1''}%
\providecommand \bibnamefont  [1]{#1}%
\providecommand \bibfnamefont [1]{#1}%
\providecommand \citenamefont [1]{#1}%
\providecommand \href@noop [0]{\@secondoftwo}%
\providecommand \href [0]{\begingroup \@sanitize@url \@href}%
\providecommand \@href[1]{\@@startlink{#1}\@@href}%
\providecommand \@@href[1]{\endgroup#1\@@endlink}%
\providecommand \@sanitize@url [0]{\catcode `\\12\catcode `\$12\catcode
  `\&12\catcode `\#12\catcode `\^12\catcode `\_12\catcode `\%12\relax}%
\providecommand \@@startlink[1]{}%
\providecommand \@@endlink[0]{}%
\providecommand \url  [0]{\begingroup\@sanitize@url \@url }%
\providecommand \@url [1]{\endgroup\@href {#1}{\urlprefix }}%
\providecommand \urlprefix  [0]{URL }%
\providecommand \Eprint [0]{\href }%
\providecommand \doibase [0]{http://dx.doi.org/}%
\providecommand \selectlanguage [0]{\@gobble}%
\providecommand \bibinfo  [0]{\@secondoftwo}%
\providecommand \bibfield  [0]{\@secondoftwo}%
\providecommand \translation [1]{[#1]}%
\providecommand \BibitemOpen [0]{}%
\providecommand \bibitemStop [0]{}%
\providecommand \bibitemNoStop [0]{.\EOS\space}%
\providecommand \EOS [0]{\spacefactor3000\relax}%
\providecommand \BibitemShut  [1]{\csname bibitem#1\endcsname}%
\let\auto@bib@innerbib\@empty
\bibitem [{\citenamefont {Ullom}\ and\ \citenamefont
  {Bennett}(2015)}]{ullomReviewSuperconductingTransitionedge2015a}%
  \BibitemOpen
  \bibfield  {author} {\bibinfo {author} {\bibfnamefont {J.~N.}\ \bibnamefont
  {Ullom}}\ and\ \bibinfo {author} {\bibfnamefont {D.~A.}\ \bibnamefont
  {Bennett}},\ }\href {\doibase 10.1088/0953-2048/28/8/084003} {\bibfield
  {journal} {\bibinfo  {journal} {Supercond. Sci. Technol.}\ }\textbf {\bibinfo
  {volume} {28}},\ \bibinfo {pages} {084003} (\bibinfo {year}
  {2015})}\BibitemShut {NoStop}%
\bibitem [{\citenamefont {Smith}\ \emph {et~al.}(2021)\citenamefont {Smith},
  \citenamefont {Adams}, \citenamefont {Bandler}, \citenamefont {Beaumont},
  \citenamefont {Chervenak}, \citenamefont {Denison}, \citenamefont {Doriese},
  \citenamefont {Durkin}, \citenamefont {Finkbeiner}, \citenamefont {Fowler},
  \citenamefont {Hilton}, \citenamefont {Hummatov}, \citenamefont {Irwin},
  \citenamefont {Kelley}, \citenamefont {Kilbourne}, \citenamefont
  {Leutenegger}, \citenamefont {Miniussi}, \citenamefont {Porter},
  \citenamefont {Reintsema}, \citenamefont {Sadleir}, \citenamefont {Sakai},
  \citenamefont {Swetz}, \citenamefont {Ullom}, \citenamefont {Vale},
  \citenamefont {Wakeham}, \citenamefont {Wassell},\ and\ \citenamefont
  {Witthoeft}}]{smithPerformanceBroadBandHighResolution2021}%
  \BibitemOpen
  \bibfield  {author} {\bibinfo {author} {\bibfnamefont {S.~J.}\ \bibnamefont
  {Smith}}, \bibinfo {author} {\bibfnamefont {J.~S.}\ \bibnamefont {Adams}},
  \bibinfo {author} {\bibfnamefont {S.~R.}\ \bibnamefont {Bandler}}, \bibinfo
  {author} {\bibfnamefont {S.}~\bibnamefont {Beaumont}}, \bibinfo {author}
  {\bibfnamefont {J.~A.}\ \bibnamefont {Chervenak}}, \bibinfo {author}
  {\bibfnamefont {E.~V.}\ \bibnamefont {Denison}}, \bibinfo {author}
  {\bibfnamefont {W.~B.}\ \bibnamefont {Doriese}}, \bibinfo {author}
  {\bibfnamefont {M.}~\bibnamefont {Durkin}}, \bibinfo {author} {\bibfnamefont
  {F.~M.}\ \bibnamefont {Finkbeiner}}, \bibinfo {author} {\bibfnamefont
  {J.~W.}\ \bibnamefont {Fowler}}, \bibinfo {author} {\bibfnamefont {G.~C.}\
  \bibnamefont {Hilton}}, \bibinfo {author} {\bibfnamefont {R.}~\bibnamefont
  {Hummatov}}, \bibinfo {author} {\bibfnamefont {K.~D.}\ \bibnamefont {Irwin}},
  \bibinfo {author} {\bibfnamefont {R.~L.}\ \bibnamefont {Kelley}}, \bibinfo
  {author} {\bibfnamefont {C.~A.}\ \bibnamefont {Kilbourne}} and others,\
  }\href {\doibase 10.1109/TASC.2021.3061918} {\bibfield  {journal} {\bibinfo
  {journal} {IEEE Transactions on Applied Superconductivity}\ }\textbf
  {\bibinfo {volume} {31}},\ \bibinfo {pages} {1} (\bibinfo {year}
  {2021})}\BibitemShut {NoStop}%
\bibitem [{\citenamefont {Sobrin}\ \emph {et~al.}(2022)\citenamefont {Sobrin},
  \citenamefont {Anderson}, \citenamefont {Bender}, \citenamefont {Benson},
  \citenamefont {Dutcher}, \citenamefont {Foster}, \citenamefont
  {{Goeckner-Wald}}, \citenamefont {Montgomery}, \citenamefont {Nadolski},
  \citenamefont {Rahlin}, \citenamefont {Ade}, \citenamefont {Ahmed},
  \citenamefont {Anderes}, \citenamefont {Archipley}, \citenamefont
  {Austermann}, \citenamefont {Avva}, \citenamefont {Aylor}, \citenamefont
  {Balkenhol}, \citenamefont {Barry}, \citenamefont {Thakur}, \citenamefont
  {Benabed}, \citenamefont {Bianchini}, \citenamefont {Bleem}, \citenamefont
  {Bouchet}, \citenamefont {Bryant}, \citenamefont {Byrum}, \citenamefont
  {Carlstrom}, \citenamefont {Carter}, \citenamefont {Cecil}, \citenamefont
  {Chang}, \citenamefont {Chaubal}, \citenamefont {Chen}, \citenamefont {Cho},
  \citenamefont {Chou}, \citenamefont {Cliche}, \citenamefont {Crawford},
  \citenamefont {Cukierman}, \citenamefont {Daley}, \citenamefont {de~Haan},
  \citenamefont {Denison}, \citenamefont {Dibert}, \citenamefont {Ding},
  \citenamefont {Dobbs}, \citenamefont {Everett}, \citenamefont {Feng},
  \citenamefont {Ferguson}, \citenamefont {Fu}, \citenamefont {Galli},
  \citenamefont {Gambrel}, \citenamefont {Gardner}, \citenamefont {Gualtieri},
  \citenamefont {Guns}, \citenamefont {Gupta}, \citenamefont {Guyser},
  \citenamefont {Halverson}, \citenamefont {{Harke-Hosemann}}, \citenamefont
  {Harrington}, \citenamefont {Henning}, \citenamefont {Hilton}, \citenamefont
  {Hivon}, \citenamefont {Holder}, \citenamefont {Holzapfel}, \citenamefont
  {Hood}, \citenamefont {Howe}, \citenamefont {Huang}, \citenamefont {Irwin},
  \citenamefont {Jeong}, \citenamefont {Jonas}, \citenamefont {Jones},
  \citenamefont {Khaire}, \citenamefont {Knox}, \citenamefont {Kofman},
  \citenamefont {Korman}, \citenamefont {Kubik}, \citenamefont {Kuhlmann},
  \citenamefont {Kuo}, \citenamefont {Lee}, \citenamefont {Leitch},
  \citenamefont {Lowitz}, \citenamefont {Lu}, \citenamefont {Meyer},
  \citenamefont {Michalik}, \citenamefont {Millea}, \citenamefont {Natoli},
  \citenamefont {Nguyen}, \citenamefont {Noble}, \citenamefont {Novosad},
  \citenamefont {Omori}, \citenamefont {Padin}, \citenamefont {Pan},
  \citenamefont {Paschos}, \citenamefont {Pearson}, \citenamefont {Posada},
  \citenamefont {Prabhu}, \citenamefont {Quan}, \citenamefont {Reichardt},
  \citenamefont {Riebel}, \citenamefont {Riedel}, \citenamefont {Rouble},
  \citenamefont {Ruhl}, \citenamefont {Saliwanchik}, \citenamefont {Sayre},
  \citenamefont {Schiappucci}, \citenamefont {Shirokoff}, \citenamefont
  {Smecher}, \citenamefont {Stark}, \citenamefont {Stephen}, \citenamefont
  {Story}, \citenamefont {Suzuki}, \citenamefont {Tandoi}, \citenamefont
  {Thompson}, \citenamefont {Thorne}, \citenamefont {Tucker}, \citenamefont
  {Umilta}, \citenamefont {Vale}, \citenamefont {Vanderlinde}, \citenamefont
  {Vieira}, \citenamefont {Wang}, \citenamefont {Whitehorn}, \citenamefont
  {Wu}, \citenamefont {Yefremenko}, \citenamefont {Yoon},\ and\ \citenamefont
  {Young}}]{sobrinDesignIntegratedPerformance2022a}%
  \BibitemOpen
  \bibfield  {author} {\bibinfo {author} {\bibfnamefont {J.~A.}\ \bibnamefont
  {Sobrin}}, \bibinfo {author} {\bibfnamefont {A.~J.}\ \bibnamefont
  {Anderson}}, \bibinfo {author} {\bibfnamefont {A.~N.}\ \bibnamefont
  {Bender}}, \bibinfo {author} {\bibfnamefont {B.~A.}\ \bibnamefont {Benson}},
  \bibinfo {author} {\bibfnamefont {D.}~\bibnamefont {Dutcher}}, \bibinfo
  {author} {\bibfnamefont {A.}~\bibnamefont {Foster}}, \bibinfo {author}
  {\bibfnamefont {N.}~\bibnamefont {{Goeckner-Wald}}}, \bibinfo {author}
  {\bibfnamefont {J.}~\bibnamefont {Montgomery}}, \bibinfo {author}
  {\bibfnamefont {A.}~\bibnamefont {Nadolski}}, \bibinfo {author}
  {\bibfnamefont {A.}~\bibnamefont {Rahlin}}, \bibinfo {author} {\bibfnamefont
  {P.~A.~R.}\ \bibnamefont {Ade}}, \bibinfo {author} {\bibfnamefont
  {Z.}~\bibnamefont {Ahmed}}, \bibinfo {author} {\bibfnamefont
  {E.}~\bibnamefont {Anderes}}, \bibinfo {author} {\bibfnamefont
  {M.}~\bibnamefont {Archipley}}, \bibinfo {author} {\bibfnamefont {J.~E.}\
  \bibnamefont {Austermann}} and others,\ }\href {\doibase
  10.3847/1538-4365/ac374f} {\bibfield  {journal} {\bibinfo  {journal} {ApJS}\
  }\textbf {\bibinfo {volume} {258}},\ \bibinfo {pages} {42} (\bibinfo {year}
  {2022})}\BibitemShut {NoStop}%
\bibitem [{\citenamefont {Hazumi}\ \emph {et~al.}(2019)\citenamefont {Hazumi},
  \citenamefont {Ade}, \citenamefont {Akiba}, \citenamefont {Alonso},
  \citenamefont {Arnold}, \citenamefont {Aumont}, \citenamefont {Baccigalupi},
  \citenamefont {Barron}, \citenamefont {Basak}, \citenamefont {Beckman},
  \citenamefont {Borrill}, \citenamefont {Boulanger}, \citenamefont {Bucher},
  \citenamefont {Calabrese}, \citenamefont {Chinone}, \citenamefont {Cho},
  \citenamefont {Cukierman}, \citenamefont {Curtis}, \citenamefont {{de Haan}},
  \citenamefont {Dobbs}, \citenamefont {Dominjon}, \citenamefont {Dotani},
  \citenamefont {Duband}, \citenamefont {Ducout}, \citenamefont {Dunkley},
  \citenamefont {Duval}, \citenamefont {Elleflot}, \citenamefont {Eriksen},
  \citenamefont {Errard}, \citenamefont {Fischer}, \citenamefont {Fujino},
  \citenamefont {Funaki}, \citenamefont {Fuskeland}, \citenamefont {Ganga},
  \citenamefont {{Goeckner-Wald}}, \citenamefont {Grain}, \citenamefont
  {Halverson}, \citenamefont {Hamada}, \citenamefont {Hasebe}, \citenamefont
  {Hasegawa}, \citenamefont {Hattori}, \citenamefont {Hattori}, \citenamefont
  {Hayes}, \citenamefont {Hidehira}, \citenamefont {Hill}, \citenamefont
  {Hilton}, \citenamefont {Hubmayr}, \citenamefont {Ichiki}, \citenamefont
  {Iida}, \citenamefont {Imada}, \citenamefont {Inoue}, \citenamefont {Inoue},
  \citenamefont {Irwin}, \citenamefont {Ishino}, \citenamefont {Jeong},
  \citenamefont {Kanai}, \citenamefont {Kaneko}, \citenamefont {Kashima},
  \citenamefont {Katayama}, \citenamefont {Kawasaki}, \citenamefont
  {Kernasovskiy}, \citenamefont {Keskitalo}, \citenamefont {Kibayashi},
  \citenamefont {Kida}, \citenamefont {Kimura}, \citenamefont {Kisner},
  \citenamefont {Kohri}, \citenamefont {Komatsu}, \citenamefont {Komatsu},
  \citenamefont {Kuo}, \citenamefont {Kurinsky}, \citenamefont {Kusaka},
  \citenamefont {Lazarian}, \citenamefont {Lee}, \citenamefont {Li},
  \citenamefont {Linder}, \citenamefont {Maffei}, \citenamefont {Mangilli},
  \citenamefont {Maki}, \citenamefont {Matsumura}, \citenamefont {Matsuura},
  \citenamefont {Meilhan}, \citenamefont {Mima}, \citenamefont {Minami},
  \citenamefont {Mitsuda}, \citenamefont {Montier}, \citenamefont {Nagai},
  \citenamefont {Nagasaki}, \citenamefont {Nagata}, \citenamefont {Nakajima},
  \citenamefont {Nakamura}, \citenamefont {Namikawa}, \citenamefont {Naruse},
  \citenamefont {Nishino}, \citenamefont {Nitta}, \citenamefont {Noguchi},
  \citenamefont {Ogawa}, \citenamefont {Oguri}, \citenamefont {Okada},
  \citenamefont {Okamoto}, \citenamefont {Okamura}, \citenamefont {Otani},
  \citenamefont {Patanchon}, \citenamefont {Pisano}, \citenamefont {Rebeiz},
  \citenamefont {Remazeilles}, \citenamefont {Richards}, \citenamefont {Sakai},
  \citenamefont {Sakurai}, \citenamefont {Sato}, \citenamefont {Sato},
  \citenamefont {Sawada}, \citenamefont {Segawa}, \citenamefont {Sekimoto},
  \citenamefont {Seljak}, \citenamefont {Sherwin}, \citenamefont {Shimizu},
  \citenamefont {Shinozaki}, \citenamefont {Stompor}, \citenamefont {Sugai},
  \citenamefont {Sugita}, \citenamefont {Suzuki}, \citenamefont {Suzuki},
  \citenamefont {Tajima}, \citenamefont {Takada}, \citenamefont {Takaku},
  \citenamefont {Takakura}, \citenamefont {Takatori}, \citenamefont {Tanabe},
  \citenamefont {Taylor}, \citenamefont {Thompson}, \citenamefont {Thorne},
  \citenamefont {Tomaru}, \citenamefont {Tomida}, \citenamefont {Tomita},
  \citenamefont {Tristram}, \citenamefont {Tucker}, \citenamefont {Turin},
  \citenamefont {Tsujimoto}, \citenamefont {Uozumi}, \citenamefont
  {Utsunomiya}, \citenamefont {Uzawa}, \citenamefont {Vansyngel}, \citenamefont
  {Wehus}, \citenamefont {Westbrook}, \citenamefont {Willer}, \citenamefont
  {Whitehorn}, \citenamefont {Yamada}, \citenamefont {Yamamoto}, \citenamefont
  {Yamasaki}, \citenamefont {Yamashita},\ and\ \citenamefont
  {Yoshida}}]{hazumiLiteBIRDSatelliteStudies2019}%
  \BibitemOpen
  \bibfield  {author} {\bibinfo {author} {\bibfnamefont {M.}~\bibnamefont
  {Hazumi}}, \bibinfo {author} {\bibfnamefont {P.~A.~R.}\ \bibnamefont {Ade}},
  \bibinfo {author} {\bibfnamefont {Y.}~\bibnamefont {Akiba}}, \bibinfo
  {author} {\bibfnamefont {D.}~\bibnamefont {Alonso}}, \bibinfo {author}
  {\bibfnamefont {K.}~\bibnamefont {Arnold}}, \bibinfo {author} {\bibfnamefont
  {J.}~\bibnamefont {Aumont}}, \bibinfo {author} {\bibfnamefont
  {C.}~\bibnamefont {Baccigalupi}}, \bibinfo {author} {\bibfnamefont
  {D.}~\bibnamefont {Barron}}, \bibinfo {author} {\bibfnamefont
  {S.}~\bibnamefont {Basak}}, \bibinfo {author} {\bibfnamefont
  {S.}~\bibnamefont {Beckman}}, \bibinfo {author} {\bibfnamefont
  {J.}~\bibnamefont {Borrill}}, \bibinfo {author} {\bibfnamefont
  {F.}~\bibnamefont {Boulanger}}, \bibinfo {author} {\bibfnamefont
  {M.}~\bibnamefont {Bucher}}, \bibinfo {author} {\bibfnamefont
  {E.}~\bibnamefont {Calabrese}}, \bibinfo {author} {\bibfnamefont
  {Y.}~\bibnamefont {Chinone}} and others,\ }\href {\doibase
  10.1007/s10909-019-02150-5} {\bibfield  {journal} {\bibinfo  {journal} {J Low
  Temp Phys}\ }\textbf {\bibinfo {volume} {194}},\ \bibinfo {pages} {443}
  (\bibinfo {year} {2019})}\BibitemShut {NoStop}%
\bibitem [{\citenamefont {Strauss}\ \emph {et~al.}(2017)\citenamefont
  {Strauss}, \citenamefont {Angloher}, \citenamefont {Bauer}, \citenamefont
  {Defay}, \citenamefont {Erb}, \citenamefont {v.~Feilitzsch}, \citenamefont
  {Iachellini}, \citenamefont {Hampf}, \citenamefont {Hauff}, \citenamefont
  {Kiefer}, \citenamefont {Lanfranchi}, \citenamefont {Langenk{\"a}mper},
  \citenamefont {Mondragon}, \citenamefont {M{\"u}nster}, \citenamefont
  {Oppenheimer}, \citenamefont {Petricca}, \citenamefont {Potzel},
  \citenamefont {Pr{\"o}bst}, \citenamefont {Reindl}, \citenamefont {Rothe},
  \citenamefont {Sch{\"o}nert}, \citenamefont {Seidel}, \citenamefont
  {Steiger}, \citenamefont {Stodolsky}, \citenamefont {Tanzke}, \citenamefont
  {Thi}, \citenamefont {Ulrich}, \citenamefont {Wawoczny}, \citenamefont
  {Willers}, \citenamefont {W{\"u}strich},\ and\ \citenamefont
  {Z{\"o}ller}}]{straussPrototypeDetectorCRESSTIII2017}%
  \BibitemOpen
  \bibfield  {author} {\bibinfo {author} {\bibfnamefont {R.}~\bibnamefont
  {Strauss}}, \bibinfo {author} {\bibfnamefont {G.}~\bibnamefont {Angloher}},
  \bibinfo {author} {\bibfnamefont {P.}~\bibnamefont {Bauer}}, \bibinfo
  {author} {\bibfnamefont {X.}~\bibnamefont {Defay}}, \bibinfo {author}
  {\bibfnamefont {A.}~\bibnamefont {Erb}}, \bibinfo {author} {\bibfnamefont
  {F.}~\bibnamefont {v.~Feilitzsch}}, \bibinfo {author} {\bibfnamefont {N.~F.}\
  \bibnamefont {Iachellini}}, \bibinfo {author} {\bibfnamefont
  {R.}~\bibnamefont {Hampf}}, \bibinfo {author} {\bibfnamefont
  {D.}~\bibnamefont {Hauff}}, \bibinfo {author} {\bibfnamefont
  {M.}~\bibnamefont {Kiefer}}, \bibinfo {author} {\bibfnamefont {J.~C.}\
  \bibnamefont {Lanfranchi}}, \bibinfo {author} {\bibfnamefont
  {A.}~\bibnamefont {Langenk{\"a}mper}}, \bibinfo {author} {\bibfnamefont
  {E.}~\bibnamefont {Mondragon}}, \bibinfo {author} {\bibfnamefont
  {A.}~\bibnamefont {M{\"u}nster}}, \bibinfo {author} {\bibfnamefont
  {C.}~\bibnamefont {Oppenheimer}} and others,\ }\href {\doibase
  10.1016/j.nima.2016.06.060} {\bibfield  {journal} {\bibinfo  {journal}
  {Nuclear Instruments and Methods in Physics Research Section A: Accelerators,
  Spectrometers, Detectors and Associated Equipment}\ }\bibinfo {series}
  {Proceedings of the {{Vienna Conference}} on {{Instrumentation}} 2016},\
  \textbf {\bibinfo {volume} {845}},\ \bibinfo {pages} {414} (\bibinfo {year}
  {2017})}\BibitemShut {NoStop}%
\bibitem [{\citenamefont {Rau}\ and\ \citenamefont
  {Collaboration}(2020)}]{rauSuperCDMSSNOLABStatus2020}%
  \BibitemOpen
  \bibfield  {author} {\bibinfo {author} {\bibfnamefont {W.}~\bibnamefont
  {Rau}}\ and\ \bibinfo {author} {\bibfnamefont {f.~S.}\ \bibnamefont
  {Collaboration}},\ }\href {\doibase 10.1088/1742-6596/1342/1/012077}
  {\bibfield  {journal} {\bibinfo  {journal} {J. Phys.: Conf. Ser.}\ }\textbf
  {\bibinfo {volume} {1342}},\ \bibinfo {pages} {012077} (\bibinfo {year}
  {2020})}\BibitemShut {NoStop}%
\bibitem [{\citenamefont {De~Fanis}\ \emph {et~al.}(2022)\citenamefont
  {De~Fanis}, \citenamefont {Ilchen}, \citenamefont {Achner}, \citenamefont
  {Baumann}, \citenamefont {Boll}, \citenamefont {Buck}, \citenamefont
  {Danilevsky}, \citenamefont {Esenov}, \citenamefont {Erk}, \citenamefont
  {Grychtol}, \citenamefont {Hartmann}, \citenamefont {Liu}, \citenamefont
  {Mazza}, \citenamefont {Monta{\~{n}}o}, \citenamefont {Music}, \citenamefont
  {Ovcharenko}, \citenamefont {Rennhack}, \citenamefont {Rivas}, \citenamefont
  {Rolles}, \citenamefont {Schmidt}, \citenamefont {Sotoudi~Namin},
  \citenamefont {Scholz}, \citenamefont {Viefhaus}, \citenamefont {Walter},
  \citenamefont {Zi{\'{o}}{\l}kowski}, \citenamefont {Zhang},\ and\
  \citenamefont {Meyer}}]{DeFanis}%
  \BibitemOpen
  \bibfield  {author} {\bibinfo {author} {\bibfnamefont {A.}~\bibnamefont
  {De~Fanis}}, \bibinfo {author} {\bibfnamefont {M.}~\bibnamefont {Ilchen}},
  \bibinfo {author} {\bibfnamefont {A.}~\bibnamefont {Achner}}, \bibinfo
  {author} {\bibfnamefont {T.~M.}\ \bibnamefont {Baumann}}, \bibinfo {author}
  {\bibfnamefont {R.}~\bibnamefont {Boll}}, \bibinfo {author} {\bibfnamefont
  {J.}~\bibnamefont {Buck}}, \bibinfo {author} {\bibfnamefont {C.}~\bibnamefont
  {Danilevsky}}, \bibinfo {author} {\bibfnamefont {S.}~\bibnamefont {Esenov}},
  \bibinfo {author} {\bibfnamefont {B.}~\bibnamefont {Erk}}, \bibinfo {author}
  {\bibfnamefont {P.}~\bibnamefont {Grychtol}}, \bibinfo {author}
  {\bibfnamefont {G.}~\bibnamefont {Hartmann}}, \bibinfo {author}
  {\bibfnamefont {J.}~\bibnamefont {Liu}}, \bibinfo {author} {\bibfnamefont
  {T.}~\bibnamefont {Mazza}}, \bibinfo {author} {\bibfnamefont
  {J.}~\bibnamefont {Monta{\~{n}}o}}, \bibinfo {author} {\bibfnamefont
  {V.}~\bibnamefont {Music}} and others,\ }\href {\doibase
  10.1107/S1600577522002284} {\bibfield  {journal} {\bibinfo  {journal}
  {Journal of Synchrotron Radiation}\ }\textbf {\bibinfo {volume} {29}},\
  \bibinfo {pages} {755} (\bibinfo {year} {2022})}\BibitemShut {NoStop}%
\bibitem [{\citenamefont {Smith}\ \emph {et~al.}(2020)\citenamefont {Smith},
  \citenamefont {Adams}, \citenamefont {Bandler}, \citenamefont {Beaumont},
  \citenamefont {Chervenak}, \citenamefont {Datesman}, \citenamefont
  {Finkbeiner}, \citenamefont {Hummatov}, \citenamefont {Kelly}, \citenamefont
  {Kilbourne}, \citenamefont {Miniussi}, \citenamefont {Porter}, \citenamefont
  {Sadleir}, \citenamefont {Sakai}, \citenamefont {Wakeham}, \citenamefont
  {Wassell}, \citenamefont {Witthoeft},\ and\ \citenamefont
  {Ryu}}]{smith100000PixelMicrocalorimeter2020}%
  \BibitemOpen
  \bibfield  {author} {\bibinfo {author} {\bibfnamefont {S.~J.}\ \bibnamefont
  {Smith}}, \bibinfo {author} {\bibfnamefont {J.~S.}\ \bibnamefont {Adams}},
  \bibinfo {author} {\bibfnamefont {S.~R.}\ \bibnamefont {Bandler}}, \bibinfo
  {author} {\bibfnamefont {S.}~\bibnamefont {Beaumont}}, \bibinfo {author}
  {\bibfnamefont {J.~A.}\ \bibnamefont {Chervenak}}, \bibinfo {author}
  {\bibfnamefont {A.~M.}\ \bibnamefont {Datesman}}, \bibinfo {author}
  {\bibfnamefont {F.~M.}\ \bibnamefont {Finkbeiner}}, \bibinfo {author}
  {\bibfnamefont {R.}~\bibnamefont {Hummatov}}, \bibinfo {author}
  {\bibfnamefont {R.~L.}\ \bibnamefont {Kelly}}, \bibinfo {author}
  {\bibfnamefont {C.~A.}\ \bibnamefont {Kilbourne}}, \bibinfo {author}
  {\bibfnamefont {A.~R.}\ \bibnamefont {Miniussi}}, \bibinfo {author}
  {\bibfnamefont {F.~S.}\ \bibnamefont {Porter}}, \bibinfo {author}
  {\bibfnamefont {J.~E.}\ \bibnamefont {Sadleir}}, \bibinfo {author}
  {\bibfnamefont {K.}~\bibnamefont {Sakai}}, \bibinfo {author} {\bibfnamefont
  {N.~A.}\ \bibnamefont {Wakeham}} and others,\ }\href {\doibase
  10.1007/s10909-020-02362-0} {\bibfield  {journal} {\bibinfo  {journal} {J Low
  Temp Phys}\ }\textbf {\bibinfo {volume} {199}},\ \bibinfo {pages} {330}
  (\bibinfo {year} {2020})}\BibitemShut {NoStop}%
\bibitem [{\citenamefont {Stevens}\ \emph {et~al.}(2020)\citenamefont
  {Stevens}, \citenamefont {Cothard}, \citenamefont {Vavagiakis}, \citenamefont
  {Ali}, \citenamefont {Arnold}, \citenamefont {Austermann}, \citenamefont
  {Choi}, \citenamefont {Dober}, \citenamefont {Duell}, \citenamefont {Duff},
  \citenamefont {Hilton}, \citenamefont {Ho}, \citenamefont {Hoang},
  \citenamefont {Hubmayr}, \citenamefont {Lee}, \citenamefont {Mangu},
  \citenamefont {Nati}, \citenamefont {Niemack}, \citenamefont {Raum},
  \citenamefont {Renzullo}, \citenamefont {Salatino}, \citenamefont {Sasse},
  \citenamefont {Simon}, \citenamefont {Staggs}, \citenamefont {Suzuki},
  \citenamefont {Truitt}, \citenamefont {Ullom}, \citenamefont {Vivalda},
  \citenamefont {Vissers}, \citenamefont {Walker}, \citenamefont {Westbrook},
  \citenamefont {Wollack}, \citenamefont {Xu},\ and\ \citenamefont
  {Yohannes}}]{stevensCharacterizationTransitionEdge2020}%
  \BibitemOpen
  \bibfield  {author} {\bibinfo {author} {\bibfnamefont {J.~R.}\ \bibnamefont
  {Stevens}}, \bibinfo {author} {\bibfnamefont {N.~F.}\ \bibnamefont
  {Cothard}}, \bibinfo {author} {\bibfnamefont {E.~M.}\ \bibnamefont
  {Vavagiakis}}, \bibinfo {author} {\bibfnamefont {A.}~\bibnamefont {Ali}},
  \bibinfo {author} {\bibfnamefont {K.}~\bibnamefont {Arnold}}, \bibinfo
  {author} {\bibfnamefont {J.~E.}\ \bibnamefont {Austermann}}, \bibinfo
  {author} {\bibfnamefont {S.~K.}\ \bibnamefont {Choi}}, \bibinfo {author}
  {\bibfnamefont {B.~J.}\ \bibnamefont {Dober}}, \bibinfo {author}
  {\bibfnamefont {C.}~\bibnamefont {Duell}}, \bibinfo {author} {\bibfnamefont
  {S.~M.}\ \bibnamefont {Duff}}, \bibinfo {author} {\bibfnamefont {G.~C.}\
  \bibnamefont {Hilton}}, \bibinfo {author} {\bibfnamefont {S.-P.~P.}\
  \bibnamefont {Ho}}, \bibinfo {author} {\bibfnamefont {T.~D.}\ \bibnamefont
  {Hoang}}, \bibinfo {author} {\bibfnamefont {J.}~\bibnamefont {Hubmayr}},
  \bibinfo {author} {\bibfnamefont {A.~T.}\ \bibnamefont {Lee}} and others,\
  }\href {\doibase 10.1007/s10909-020-02375-9} {\bibfield  {journal} {\bibinfo
  {journal} {J Low Temp Phys}\ }\textbf {\bibinfo {volume} {199}},\ \bibinfo
  {pages} {672} (\bibinfo {year} {2020})},\ \Eprint
  {http://arxiv.org/abs/1912.00860} {1912.00860} \BibitemShut {NoStop}%
\bibitem [{\citenamefont {Gottardi}\ and\ \citenamefont
  {Nagayashi}(2021)}]{gottardiReviewXrayMicrocalorimeters2021}%
  \BibitemOpen
  \bibfield  {author} {\bibinfo {author} {\bibfnamefont {L.}~\bibnamefont
  {Gottardi}}\ and\ \bibinfo {author} {\bibfnamefont {K.}~\bibnamefont
  {Nagayashi}},\ }\href {\doibase 10.3390/app11093793} {\bibfield  {journal}
  {\bibinfo  {journal} {Applied Sciences}\ }\textbf {\bibinfo {volume} {11}},\
  \bibinfo {pages} {3793} (\bibinfo {year} {2021})}\BibitemShut {NoStop}%
\bibitem [{\citenamefont {Arion}\ and\ \citenamefont
  {Hergenhahn}(2015)}]{arionCoincidenceSpectroscopyPresent2015}%
  \BibitemOpen
  \bibfield  {author} {\bibinfo {author} {\bibfnamefont {T.}~\bibnamefont
  {Arion}}\ and\ \bibinfo {author} {\bibfnamefont {U.}~\bibnamefont
  {Hergenhahn}},\ }\href {\doibase 10.1016/j.elspec.2015.06.004} {\bibfield
  {journal} {\bibinfo  {journal} {Journal of Electron Spectroscopy and Related
  Phenomena}\ }\bibinfo {series} {Special {{Anniversary Issue}}: {{Volume}}
  200},\ \textbf {\bibinfo {volume} {200}},\ \bibinfo {pages} {222} (\bibinfo
  {year} {2015})}\BibitemShut {NoStop}%
\bibitem [{\citenamefont {Patel}\ \emph {et~al.}(2021)\citenamefont {Patel},
  \citenamefont {Withington}, \citenamefont {Thomas}, \citenamefont {Goldie},\
  and\ \citenamefont {Shard}}]{patelSimulationMethodInvestigating2021a}%
  \BibitemOpen
  \bibfield  {author} {\bibinfo {author} {\bibfnamefont {K.~M.}\ \bibnamefont
  {Patel}}, \bibinfo {author} {\bibfnamefont {S.}~\bibnamefont {Withington}},
  \bibinfo {author} {\bibfnamefont {C.~N.}\ \bibnamefont {Thomas}}, \bibinfo
  {author} {\bibfnamefont {D.~J.}\ \bibnamefont {Goldie}}, \ and\ \bibinfo
  {author} {\bibfnamefont {A.~G.}\ \bibnamefont {Shard}},\ }\href {\doibase
  10.1088/1361-6668/ac30d0} {\bibfield  {journal} {\bibinfo  {journal}
  {Supercond. Sci. Technol.}\ }\textbf {\bibinfo {volume} {34}},\ \bibinfo
  {pages} {125007} (\bibinfo {year} {2021})}\BibitemShut {NoStop}%
\bibitem [{\citenamefont
  {Harwin}(2021)}]{harwinMicroscopicPhysicsTransition2021}%
  \BibitemOpen
  \bibfield  {author} {\bibinfo {author} {\bibfnamefont {R.}~\bibnamefont
  {Harwin}},\ }\emph {\bibinfo {title} {Microscopic Physics of Transition Edge
  Sensors}},\ \href {\doibase 10.17863/CAM.64460} {\bibinfo {type} {Thesis}},\
  \bibinfo  {school} {University of Cambridge} (\bibinfo {year}
  {2021})\BibitemShut {NoStop}%
\bibitem [{\citenamefont {Irwin}\ and\ \citenamefont
  {Hilton}(2005)}]{irwinTransitionEdgeSensors2005}%
  \BibitemOpen
  \bibfield  {author} {\bibinfo {author} {\bibfnamefont {K.}~\bibnamefont
  {Irwin}}\ and\ \bibinfo {author} {\bibfnamefont {G.}~\bibnamefont {Hilton}},\
  }in\ \href {\doibase 10.1007/10933596_3} {\emph {\bibinfo {booktitle}
  {Cryogenic {{Particle Detection}}}}},\ \bibinfo {series and number} {Topics
  in {{Applied Physics}}},\ \bibinfo {editor} {edited by\ \bibinfo {editor}
  {\bibfnamefont {C.}~\bibnamefont {Enss}}}\ (\bibinfo  {publisher}
  {{Springer}},\ \bibinfo {address} {{Berlin, Heidelberg}},\ \bibinfo {year}
  {2005})\ pp.\ \bibinfo {pages} {63--150}\BibitemShut {NoStop}%
\bibitem [{\citenamefont
  {{Figueroa-Feliciano}}(2001)}]{figueroa-felicianoTheoryDevelopmentPositionsensitive2001}%
  \BibitemOpen
  \bibfield  {author} {\bibinfo {author} {\bibfnamefont {E.}~\bibnamefont
  {{Figueroa-Feliciano}}},\ }\emph {\bibinfo {title} {Theory and Development of
  Position-Sensitive Quantum Calorimeters}},\ \href@noop {} {Ph.D. thesis},\
  \bibinfo  {school} {Stanford University} (\bibinfo {year} {2001})\BibitemShut
  {NoStop}%
\bibitem [{\citenamefont {Moseley}, \citenamefont {Mather},\ and\ \citenamefont
  {McCammon}(1984)}]{moseleyThermalDetectorsRay1984}%
  \BibitemOpen
  \bibfield  {author} {\bibinfo {author} {\bibfnamefont {S.~H.}\ \bibnamefont
  {Moseley}}, \bibinfo {author} {\bibfnamefont {J.~C.}\ \bibnamefont {Mather}},
  \ and\ \bibinfo {author} {\bibfnamefont {D.}~\bibnamefont {McCammon}},\
  }\href {\doibase 10.1063/1.334129} {\bibfield  {journal} {\bibinfo  {journal}
  {Journal of Applied Physics}\ }\textbf {\bibinfo {volume} {56}},\ \bibinfo
  {pages} {1257} (\bibinfo {year} {1984})}\BibitemShut {NoStop}%
\bibitem [{\citenamefont {Goto}\ and\ \citenamefont
  {Takeichi}(1996)}]{gotoCarbonReferenceAuger1996a}%
  \BibitemOpen
  \bibfield  {author} {\bibinfo {author} {\bibfnamefont {K.}~\bibnamefont
  {Goto}}\ and\ \bibinfo {author} {\bibfnamefont {Y.}~\bibnamefont
  {Takeichi}},\ }\href {\doibase 10.1116/1.579962} {\bibfield  {journal}
  {\bibinfo  {journal} {Journal of Vacuum Science \& Technology A}\ }\textbf
  {\bibinfo {volume} {14}},\ \bibinfo {pages} {1408} (\bibinfo {year}
  {1996})}\BibitemShut {NoStop}%
\bibitem [{\citenamefont
  {Patel}(2023)}]{patelTransitionEdgeSensorsElectron2023}%
  \BibitemOpen
  \bibfield  {author} {\bibinfo {author} {\bibfnamefont {K.~M.}\ \bibnamefont
  {Patel}},\ }\emph {\bibinfo {title} {Transition-{{Edge Sensors}} for
  {{Electron Spectroscopy}}}},\ \href@noop {} {Ph.D. thesis},\ \bibinfo
  {school} {University of Cambridge} (\bibinfo {year} {2023})\BibitemShut
  {NoStop}%
\bibitem [{\citenamefont {Lee}\ \emph {et~al.}(2015)\citenamefont {Lee},
  \citenamefont {Adams}, \citenamefont {Bandler}, \citenamefont {Chervenak},
  \citenamefont {Eckart}, \citenamefont {Finkbeiner}, \citenamefont {Kelley},
  \citenamefont {Kilbourne}, \citenamefont {Porter}, \citenamefont {Sadleir},
  \citenamefont {Smith},\ and\ \citenamefont
  {Wassell}}]{leeFinePitchTransitionedge2015a}%
  \BibitemOpen
  \bibfield  {author} {\bibinfo {author} {\bibfnamefont {S.~J.}\ \bibnamefont
  {Lee}}, \bibinfo {author} {\bibfnamefont {J.~S.}\ \bibnamefont {Adams}},
  \bibinfo {author} {\bibfnamefont {S.~R.}\ \bibnamefont {Bandler}}, \bibinfo
  {author} {\bibfnamefont {J.~A.}\ \bibnamefont {Chervenak}}, \bibinfo {author}
  {\bibfnamefont {M.~E.}\ \bibnamefont {Eckart}}, \bibinfo {author}
  {\bibfnamefont {F.~M.}\ \bibnamefont {Finkbeiner}}, \bibinfo {author}
  {\bibfnamefont {R.~L.}\ \bibnamefont {Kelley}}, \bibinfo {author}
  {\bibfnamefont {C.~A.}\ \bibnamefont {Kilbourne}}, \bibinfo {author}
  {\bibfnamefont {F.~S.}\ \bibnamefont {Porter}}, \bibinfo {author}
  {\bibfnamefont {J.~E.}\ \bibnamefont {Sadleir}}, \bibinfo {author}
  {\bibfnamefont {S.~J.}\ \bibnamefont {Smith}}, \ and\ \bibinfo {author}
  {\bibfnamefont {E.~J.}\ \bibnamefont {Wassell}},\ }\href {\doibase
  10.1063/1.4936793} {\bibfield  {journal} {\bibinfo  {journal} {Applied
  Physics Letters}\ }\textbf {\bibinfo {volume} {107}},\ \bibinfo {pages}
  {223503} (\bibinfo {year} {2015})}\BibitemShut {NoStop}%
\bibitem [{\citenamefont {Ding}\ \emph {et~al.}(2004)\citenamefont {Ding},
  \citenamefont {Li}, \citenamefont {Goto}, \citenamefont {Jiang},\ and\
  \citenamefont {Shimizu}}]{dingEnergySpectraBackscattered2004a}%
  \BibitemOpen
  \bibfield  {author} {\bibinfo {author} {\bibfnamefont {Z.~J.}\ \bibnamefont
  {Ding}}, \bibinfo {author} {\bibfnamefont {H.~M.}\ \bibnamefont {Li}},
  \bibinfo {author} {\bibfnamefont {K.}~\bibnamefont {Goto}}, \bibinfo {author}
  {\bibfnamefont {Y.~Z.}\ \bibnamefont {Jiang}}, \ and\ \bibinfo {author}
  {\bibfnamefont {R.}~\bibnamefont {Shimizu}},\ }\href {\doibase
  10.1063/1.1791752} {\bibfield  {journal} {\bibinfo  {journal} {Journal of
  Applied Physics}\ }\textbf {\bibinfo {volume} {96}},\ \bibinfo {pages} {4598}
  (\bibinfo {year} {2004})}\BibitemShut {NoStop}%
\bibitem [{\citenamefont {Walker}\ \emph {et~al.}(2008)\citenamefont {Walker},
  \citenamefont {{El-Gomati}}, \citenamefont {Assa'd},\ and\ \citenamefont
  {Zadra{\v z}il}}]{walkerSecondaryElectronEmission2008}%
  \BibitemOpen
  \bibfield  {author} {\bibinfo {author} {\bibfnamefont {C.}~\bibnamefont
  {Walker}}, \bibinfo {author} {\bibfnamefont {M.}~\bibnamefont {{El-Gomati}}},
  \bibinfo {author} {\bibfnamefont {A.}~\bibnamefont {Assa'd}}, \ and\ \bibinfo
  {author} {\bibfnamefont {M.}~\bibnamefont {Zadra{\v z}il}},\ }\href {\doibase
  10.1002/sca.20124} {\bibfield  {journal} {\bibinfo  {journal} {Scanning}\
  }\textbf {\bibinfo {volume} {30}},\ \bibinfo {pages} {365} (\bibinfo {year}
  {2008})}\BibitemShut {NoStop}%
\bibitem [{\citenamefont {{Albert C. Thompson}}\ \emph
  {et~al.}(2009)\citenamefont {{Albert C. Thompson}}, \citenamefont {{Janos
  Kirz}}, \citenamefont {{David T. Atwood}}, \citenamefont {{Eric M.
  Gullikson}}, \citenamefont {{Malcom R. Howells}}, \citenamefont {{Jeffrey B.
  Kortright}}, \citenamefont {{Yanwei Liu}}, \citenamefont {{Arthur L.
  Robinson}}, \citenamefont {{James H. Underwood}}, \citenamefont {{Kwang-Je
  Kim}}, \citenamefont {{Ingolf Lindau}}, \citenamefont {{Piero Pianetta}},
  \citenamefont {{Herman Winick}}, \citenamefont {{Gwyn P. Williams}},\ and\
  \citenamefont {{James H. Scofield}}}]{albertc.thompsonXrayDataBooklet2009}%
  \BibitemOpen
  \bibfield  {author} {\bibinfo {author} {\bibnamefont {{Albert C. Thompson}}},
  \bibinfo {author} {\bibnamefont {{Janos Kirz}}}, \bibinfo {author}
  {\bibnamefont {{David T. Atwood}}}, \bibinfo {author} {\bibnamefont {{Eric M.
  Gullikson}}}, \bibinfo {author} {\bibnamefont {{Malcom R. Howells}}},
  \bibinfo {author} {\bibnamefont {{Jeffrey B. Kortright}}}, \bibinfo {author}
  {\bibnamefont {{Yanwei Liu}}}, \bibinfo {author} {\bibnamefont {{Arthur L.
  Robinson}}}, \bibinfo {author} {\bibnamefont {{James H. Underwood}}},
  \bibinfo {author} {\bibnamefont {{Kwang-Je Kim}}}, \bibinfo {author}
  {\bibnamefont {{Ingolf Lindau}}}, \bibinfo {author} {\bibnamefont {{Piero
  Pianetta}}}, \bibinfo {author} {\bibnamefont {{Herman Winick}}}, \bibinfo
  {author} {\bibnamefont {{Gwyn P. Williams}}}, \ and\ \bibinfo {author}
  {\bibnamefont {{James H. Scofield}}},\ }\href@noop {} {\emph {\bibinfo
  {title} {X-Ray {{Data Booklet}}}}},\ \bibinfo {edition} {third edition}\ ed.\
  (\bibinfo  {publisher} {{Lawrence Berkeley National Laboratory, University of
  California}},\ \bibinfo {year} {2009})\BibitemShut {NoStop}%
\bibitem [{\citenamefont {Patel}\ \emph {et~al.}(2023)\citenamefont {Patel},
  \citenamefont {Goldie}, \citenamefont {Withington},\ and\ \citenamefont
  {Thomas}}]{patelSensitivityTransitionedgeSensors2023}%
  \BibitemOpen
  \bibfield  {author} {\bibinfo {author} {\bibfnamefont {K.~M.}\ \bibnamefont
  {Patel}}, \bibinfo {author} {\bibfnamefont {D.~J.}\ \bibnamefont {Goldie}},
  \bibinfo {author} {\bibfnamefont {S.}~\bibnamefont {Withington}}, \ and\
  \bibinfo {author} {\bibfnamefont {C.~N.}\ \bibnamefont {Thomas}},\ }in\ \href
  {\doibase 10.1117/12.2666646} {\emph {\bibinfo {booktitle} {{{SPIE Future
  Sensing Technologies}} 2023}}},\ Vol.\ \bibinfo {volume} {12327}\ (\bibinfo
  {publisher} {{SPIE}},\ \bibinfo {year} {2023})\ pp.\ \bibinfo {pages}
  {36--41}\BibitemShut {NoStop}%
\end{thebibliography}%
\bibliographystyle{custom.bst}
\end{document}